\documentclass[amssymb,amsmath,aps,showpacs,floatfix,nofootinbib,showpacs,12pt]{revtex4}
\usepackage{amssymb}
\usepackage{graphicx}
\usepackage{color}
\usepackage{soul}
\usepackage{latexsym}

\newcommand{\lsim}{\lesssim}
\newcommand{\gsim}{\gtrsim}

\def\lsim{\mathrel{\raise.3ex\hbox{$<$\kern-.75em\lower1ex\hbox{$\sim$}}}}
\def\gsim{\mathrel{\raise.3ex\hbox{$>$\kern-.75em\lower1ex\hbox{$\sim$}}}}

\def\beq{\begin{equation}}
\def\eeq{\end{equation}}
\def\beqn{\begin{eqnarray}}
\def\eeqn{\end{eqnarray}}
\def\bea{\begin{eqnarray}}
\def\eea{\end{eqnarray}}
\def\be{\begin{equation}}
\def\ee{\end{equation}}
\newcommand{\fslash}[1]{{#1 \kern -0.7em/ \kern 0.1em}}

\begin{document}

\voffset 1.25cm

\title{ Constraints on the Dark Matter Annihilations by Neutrinos with
Substructure Effects Included}

\author{Peng-fei Yin $^{1}$, Jia Liu $^{1}$, Qiang Yuan $^{3}$, Xiao-jun Bi $^{2,3}$ and
Shou-hua Zhu $^{1}$}

\affiliation{$^{1}$ Institute of Theoretical Physics \& State Key
Laboratory of Nuclear Physics and Technology, Peking University,
Beijing 100871, China \\ $^{2}$ Center for High Energy Physics,
Peking University, Beijing 100871, China \\ $^{3}$ Key Laboratory of
Particle Astrophysics, Institute of High Energy Physics, Chinese
Academy of Sciences, Beijing 100049, P. R. China}

\date{\today}

\begin{abstract}

Dark matter (DM) annihilations in the Galaxy may produce high energy
neutrinos, which can be detected by the neutrino telescopes, for
example IceCube, ANTARES and Super-Kamiokande. The neutrinos can
also arise from hadronic interaction between cosmic ray and
atmosphere around the Earth, known as atmospheric neutrino. Current
measurements on neutrino flux is consistent with theoretical
prediction of atmospheric neutrino within the uncertainties. In this
paper, by requiring that the DM annihilation neutrino flux is less
than the current measurements, we obtain an upper bound on the cross
section of dark matter annihilation $ \left\langle {\sigma v}
\right\rangle$. Compared with previous investigations, we improve
the bound by including DM substructure contributions. In our paper,
two kinds of substructure effects are scrutinized. One is the
substructure average contribution over all directions. The other is
point source effect by single massive sub-halo. We found that the
former can improve the bound by several times, while the latter can
improve the bound by $ 10^1 \sim 10^4$ utilizing the excellent
angular resolution of neutrino telescope IceCube. The exact
improvement depends on the DM profile and the sub-halo concentration
model. In some model, IceCube can achieve the sensitivity of $
\left\langle {\sigma v} \right\rangle \sim 10^{ - 26} cm^3 s^{ - 1}
$.

\end{abstract}



\maketitle

\section{Introduction}

Many astronomical observations indicate that most of the matter in
our universe is dark (see e.g. Ref. \cite{Jungman:1995df}). The
evidences come mainly from the gravitational effects of the dark
matter (DM), such as the rotation curves of spiral galaxies
\cite{Begeman:1991iy,Persic:1995ru}, the gravitational lensing
\cite{Tyson95} and the dynamics of galaxy clusters \cite{White93}.
The studies such as primordial nucleosynthesis \cite{Peebles71} and
cosmic microwave background (CMB) \cite{Spergel03} show that the DM
is mostly non-baryonic. Combining recent cosmological measurements,
for example from the Wilkinson Microwave Anisotropy Probe (WMAP),
one could deduce precisely the relic density of DM, namely
$\Omega_{DM}h^2=0.1143\pm 0.0034$ \cite{Hinshaw:2008kr}. However,
the nature of dark matter is still unclear. In the literature there
is a ``zoo'' of particle candidates for DM \cite{Bertone04}, among
which the most popular candidate at present is the weakly
interacting massive particle (WIMP) such as the lightest
supersymmetric particle (LSP), lightest Kaluza-Klein particle (LKP)
$etc$.

Search for WIMP in particle physics experiments is very important to
pin down the properties of the DM. Besides searching missing energy
signals at accelerator-based experiments, there are usually two
classes of methods to detect WIMP, namely direct and indirect ones.
The former method detects WIMP by measuring the recoil of heavy
nucleus in the detector and gives the most strong evidence for the
existence of DM. The latter one detects the DM self-annihilation
signals, which include neutrinos, photons, anti-matter particles and
so on. Among them neutrinos are one of the most attractive signals.
For the {\em low energy} neutrinos (say much less than 100 GeV),
their interactions with matter are highly suppressed by a factor at
least $Q^2/m_W^2$ with $Q$ the typical energy scale of the
interaction. The neutrinos are hardly energy loss and trajectory
deflection during their propagation, therefore they may carry the
information of the nature and distribution of the DM. However due to
the same reason, it is hard to capture such kind of {\em low energy}
and relatively low flux neutrinos. For the {\em high energy}
neutrinos (say around 100 GeV or higher), the interactions among
neutrinos and matter become much stronger. These neutrinos may keep
the information of the DM, and it is relatively easy to observe them
experimentally.

 One proposal of detecting the high energy neutrino signals is to explore the
locations close to us such as the center of the Sun or the Earth to
get enough neutrino flux. The DM particles are gravitational trapped
in the center of the Sun or Earth and produce neutrinos by
annihilation \cite{Liu:2008kz}. If the annihilation and capture
processes are in equilibrium, the neutrino flux are mainly
determined by the cross section of the DM and nuclei. Another
proposal is to detect the neutrino signals from DM annihilation in
the Milky Way (MW). Though the sources in the MW are farther than
the Sun, it is natural to expect that the larger amount of DM can
compensate the distance. Moreover the neutrino flux depends on the
DM annihilation rate and number density square, therefore the
regions with high density in the MW, such as the Galactic Center
(GC) or sub-halos, should be potentially excellent observational
targets.

The GC is conventionally thought to be source-rich astrophysical
laboratory and has attracted many attention of astronomers. It is
also true for the DM indirect searches. Due to the weak interaction
of DM particles, the DM density at the GC is highly accumulated as
shown by detail simulations, which makes the GC a bright source of
DM annihilation. However, the complicated astrophysical environment
and various kinds of astrophysical sources  make the GC a high
background site. In addition, the overlapping with bayonic matter
objects (e.g., the central massive black hole \cite{Gondolo:1999ef})
may also affect the DM distribution and increase the uncertainties.
It should be emphasized that the galactic sub-halos may be good
candidates as DM sources. Since the self-annihilation of DM is
square-dependent on the number density, the clump of substructure is
expected to effectively enhance the annihilation signal and plays a
role of the so-called ``boost factor'' \cite{bi06a,bi06b,yuan07}.
Furthermore as indicated by simulations, the spatial distribution of
DM sub-halos is tend to be spherical symmetric in the MW halo, which
may locate at a low-background site and effectively avoid the source
confusion in the galactic plane. The effects of DM sub-halos on the
flux of induced neutrino is what we try to investigate in this work.

The neutrinos detected by high-energy neutrino telescopes such as
Super-Kamikande \cite{Desai:2004pq}, AMANDA \cite{Ahrens:2003fg}
etc. are thought to be mainly from atmospheric neutrinos. Here they
originate from the decay of hadrons which are produced by the strong
interactions of cosmic rays with atmosphere. Experimentally no
obvious excess has yet been observed. Then the measurements on
neutrino flux can be utilized to set bounds on the DM annihilation
cross section. As neutrinos are the most difficult to detect in the
DM annihilating final states, the authors of Ref.
\cite{Beacom:2006tt} and \cite{Yuksel:2007ac} assumed that the DM
annihilate solely into neutrinos. They calculated the extra-galactic
and the galactic neutrino fluxes, compared them with atmospheric
neutrino flux, and set an upper bound on the DM annihilation cross
section $\langle\sigma v\rangle$. However, the DM may annihilate
into final states other than neutrinos. In most of the DM models
there are several annihilation channels. Moreover high energy
neutrinos from the DM annihilation will lead to gauge bosons
bremsstrahlung \cite{Kachelriess:2007aj,Bell:2008ey} even in the
standard model (SM). The electromagnetic final states through
higher-order corrections are also inevitable \cite{Dent:2008qy}.
Therefore the assumption that DM annihilate into only neutrinos
gives the most conservative bound on DM annihilation cross section.

In this paper we calculated the neutrino flux from the DM
annihilations in the MW including the contributions from sub-halos
by assuming that the DM annihilate into neutrinos only
\cite{Beacom:2006tt,Yuksel:2007ac}. By comparing the predicted flux
with the available atmospheric neutrino measurements, we set a very
strict constraint on the DM total annihilation cross section.
Compared to the previous studies, in this work we utilize the
angular resolution of the neutrino telescope to derive the stricter
constraints. Here the massive sub-halos can be treated as the
point-like sources. Based on our analysis we may observe the high
energy neutrino flux provided that precise angular resolution data
from ANTARES \cite{Pradier:2008iv} and IceCube \cite{Ahrens:2003ix}
is available. On the other hand if no excess flux out of atmospheric
neutrino is observed, an improved upper-bound of the annihilation
cross section and/or the exclusion of certain sub-halo models can be
obtained.

This paper is organized as following. In Sec. II, we describe the
sub-halo models according to the N-body simulation results. In Sec.
III, we give the constraints on the dark matter annihilation cross
section. The conclusions and discussions are given in Sec. IV.

\section{Galactic DM distribution and substructure}

The current knowledge of the DM spatial distribution is mostly from
the N-body simulation. Navarro et al. \cite{nfw97} firstly proposed
a universal DM profile (referred as ``NFW''). Based on their
simulation in a wide range of halo mass, the density of DM can be
written as \cite{nfw97}
\begin{equation}
\rho(r)=\frac{\rho_s}{(r/r_s)[1+r/r_s]^2}, \label{nfw}
\end{equation}
where $\rho_s$ and $r_s$ are the density scale and radius parameters
for a specific DM halo. Moore et al. \cite{moore98} gave another
profile with a more cusped inner slope compared with NFW as
\begin{equation}
\rho(r)=\frac{\rho_s}{(r/r_s)^{1.5}[1+(r/r_s)^{1.5}]}.
\label{moore}
\end{equation}
In addition, cored profile as
$\frac{\rho_s}{(1+r/r_s)[1+(r/r_s)^2]}$
\cite{burkert95,Salucci:2000ps} or cuspy profile with different
inner slope from NFW and Moore (e.g., \cite{diemand05}) were also
proposed in the literature. Reed et al. showed that even the inner
slope steepens with the decrement of the halo mass, instead of a
universal one \cite{reed05}. All these profiles give the same
behaviors $\sim r^{-3}$ at large radii, but show discrepancies in
the inner region of the halo. Precise determination of the DM
profile needs simulation with higher resolution and better
understanding of the DM properties such as the interaction with
baryonic matter. In this work we will adopt both NFW and Moore
profiles for the discussion. It should be noted that the central
density for NFW or Moore profile is divergent. In order to handle
the singularity, a characteristic radius $r_c$ is introduced within
which the DM density is kept a constant $\rho_{max}$ due to the
balance between the annihilation rate and the in-falling rate of DM
\cite{berezinsky92}. Typically we have $\rho_{max}=
10^{18}\sim10^{19}$ M$_{\odot}$ kpc$^{-3}$ \cite{lavalle08}.

\subsection{Determination of the profile parameters $r_s$ and $\rho_s$ }

Following \cite{bullock01}, we use the virial mass $M_v$ of the halo
to determine the parameters $\rho_s$ and $r_s$. For a DM halo with
specified mass $M_v$, the virial radius $r_v$ is defined as
\begin{equation}
r_v=\left(\frac{M_v}{(4\pi/3)\Delta\rho_c}\right)^{1/3}, \label{rv}
\end{equation}
with the density amplifying factor over the background
$\Delta\approx200$ and the critical density of the universe
$\rho_c=139$ M$_{\odot}$ kpc$^{-3}$. The concentration parameter
$c_v$ (CP) is defined as

\begin{equation}
c_v=\frac{r_v}{r_{-2}},
\label{cv}
\end{equation}
where $r_{-2}$ refers to the radius at which $\frac{{\rm d} \left(
r^2\rho \right)}{dr} |_{r=r_{-2}}=0$. It is shown that for NFW
profile $r_{-2}=r_s^{nfw}$ and for Moore profile
$r_{-2}=0.63\,r_s^{moore}$, so we have
\begin{equation}
r_s^{nfw}=\frac{r_v(M_v)}{c_v(M_v)},\ \
r_s^{moore}=\frac{r_v(M_v)}{0.63\,c_v(M_v)}.
\label{rs}
\end{equation}
Then $\rho_s$ can be derived just by requiring $\int\rho(r){\rm
d}V=M_v$. We can see that the profile of the DM halo is fully
determined provided that the $c_v - M_v$ relation is specified.

Generally the $c_v-M_v$ relation is fitted from the numerical
simulation. Here we will use two toy models of Eke et al.
(\cite{eke01}, denoted by ENS01) and Bullock et al.
(\cite{bullock01}, denoted by B01), within which the DM halo forms
based on the cosmological background density field. The CP predicted
in these models increases with the decrement of the halo mass. Such
behavior is understandable in the frame of hierarchy structure
formation, i.e. smaller halo forms earlier when the universe is
denser than today. This behavior is confirmed at the cluster scale
\cite{buote07,comerford07}. However, other studies showed agreement
or disagreement with these two models, which indicate that we may
not achieve the final understanding of this topic at present (see
the discussion of Ref. \cite{lavalle08}). For the current work,
these models are regarded as reference ones. We use the fitted
polynomial form of the simulation  at $z=0$ and extrapolate to low
masses \cite{lavalle08}:
\begin{equation}
\ln(c_v)=\sum_{i=0}^4C_i\times\left[\ln\frac{M_v}{M_{\odot}}\right]^i,
\label{lncv}
\end{equation}
with $M_{\odot}$ the mass of the Sun and
\begin{equation}
C_i^{ENS01}=\{3.14,-0.018,-4.06\times10^{-4},0,0\}
\label{cvens01}
\end{equation}
and
\begin{equation}
C_i^{B01}=\{4.34,-0.0384,-3.91\times10^{-4},-2.2\times10^{-6},
-5.5\times10^{-7}\}.
\label{cvb01}
\end{equation}

\begin{figure}[!htb]
\begin{center}
\includegraphics{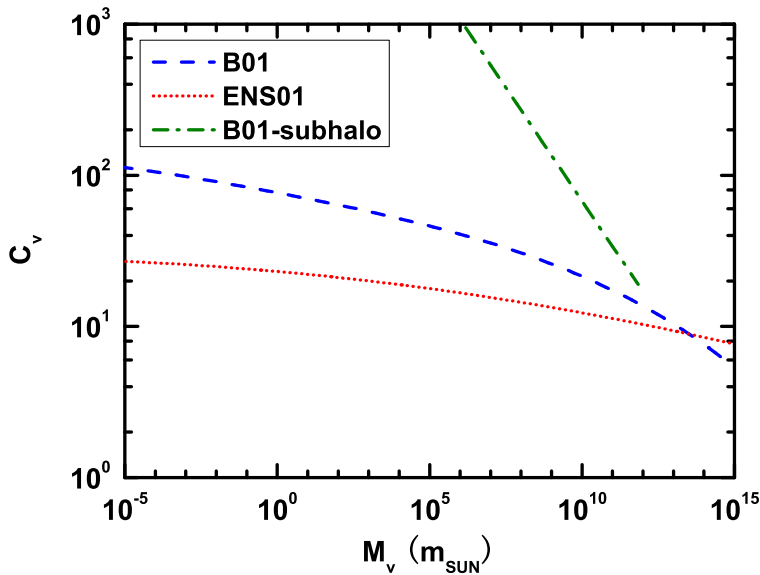}
\caption{Concentration parameter (CP)  $c_v$ as a function of halo
mass $M_v$ at epoch $z=0$. }\label{cveps}
\end{center}
\end{figure}

Fig. \ref{cveps} shows $c_v$ as a function of the halo mass $M_v$
(see also Fig. 1 of Ref. \cite{cola06}). For the very low mass, it
is shown that $c_v$ becomes flat due to the fact that small objects
tend to collapse at the same epoch. It should be noted that ENS01
and B01 models are for distinct halos in the universe. For the
sub-halos within a host halo which is denser than the universe
background, as we will discuss in this work, it is expected to be
more concentrated than the distinct halos. In Ref. \cite{bullock01}
the simulation showed the sub-halos within a host halo indeed have
larger $c_v$ than the distinct ones with the same mass. The
simulation shows the $c_v$ of subhalo has
 $c_v \sim M_v^{-0.3}$, which is steeper than the distinct
halo ($c_v \sim M_v^{-0.13}$). In Fig. \ref{cveps} we also show the
extrapolated results of $c_v$ for sub-halos with mass $10^6\sim
10^{12}$ M$_{\odot}$. The other way to deal with the sub-halo is to
multiply the result for distinct one by an empirical factor (e.g.,
$\sim2$ in Ref. \cite{cola06}, in the following we denote this model
as $B01\times2$).

\subsection{The MW halo and substructure}

The mass of the MW DM halo is about $1\sim2\times10^{12}$
M$_{\odot}$ determined from the rotation curve or kinematics of
tracer populations such as the stars, satellite galaxies and
globular clusters \cite{wilkinson99,sakamoto03,smith07}. A recent
work by Xue et al. showed that $M_v=\left( 0.93\pm0.25
\right)\times10^{12}$ M$_{\odot}$ through an analysis of kinematics
of a large sample of SDSS halo stars \cite{xue08}. Here we adopt
$M_v=10^{12}$ M$_{\odot}$ as the mass of the total MW halo. As shown
below, about $10\%\sim20\%$ of the mass will form substructures, so
the mass of the smooth halo is $0.8\sim0.9\times10^{12}$
M$_{\odot}$. A NFW profile is adopted for the smooth halo. Using the
B01 model, we find that\footnote{Here $10^{12}$ M$_{\odot}$ is used
to calculate $r_v$ and $c_v$, while the density at the solar
location is scaled to match that of the mass of the smooth component
is $0.85\times10^{12}$ M$_{\odot}$.} $r_v\approx205$ kpc,
$c_v\approx13.6$ and $\rho_{\odot}\approx0.34$ GeV cm$^{-3}$. This
configuration of the smooth halo is fixed in the following
discussion since we will focus on the substructure in the present
work.

The survival of substructure in galactic halo was revealed by many
simulation groups
\cite{tormen98,klypin99,moore99,ghigna00,springel01,
zentner03,lucia04}. A recent simulation conduced by Diemand et al.
showed that the self-bound substructure could even be as light as
the Earth, with a huge number reaching $\sim10^{15}$
\cite{diemand05}. Simulations give the number density of sub-halos
an isothermal spatial distribution and a power-law mass function as
\begin{equation}
\frac{{\rm d}N}{{\rm d}M_{sub}\cdot4\pi r^2{\rm d}r}=
N_0\left(\frac{M_{sub}}{M_{host}}\right)^{-\alpha}
\frac{1}{1+\left(\frac{r}{r_H}\right)^2}, \label{number}
\end{equation}
where $M_{sub}$ and $M_{host}$ are the masses of sub-halo and host
halo ($10^{12}$ M$_{\odot}$ here), $r_H$ is the core radius which
usually is a fraction of the virial radius of the host halo,
$\alpha$ is the slope of the mass function  and $N_0$ is the
normalization factor. For a galactic host halo, $r_H$ was found to
be about $0.14\,r_v$ \cite{diemand04}. The slope $\alpha$ lies
between $1.7$ and $2.1$ in various works \cite{moore99,
ghigna00,helmi02,gao04,lucia04,shaw06}. Here we adopt $\alpha=1.9$
as in Ref. \cite{bi06a,cola06,yuan07}. The mass function of Eq.
(\ref{number}) is thought to be held in the mass range from the
minimal sub-halo with mass $\sim10^{-6}$ M$_{\odot}$ which is close
to the free-streaming mass
\cite{hofmann01,chen01,green05,diemand05}, to the maximum one about
$0.01\,M_{host}$ \cite{bi06a}. The normalization is determined by
setting the number of sub-halos with mass larger than $10^8$
M$_{\odot}$ is 100 \cite{lavalle08}. Finally in the inner region of
the host halo, strong tidal force tends to destroy the sub-halo and
the survival number should be cut down. We employ the ``tidal
approximation'' as in Ref. \cite{bi06a}. Under this configuration,
we find that the mass fraction of substructure is about $14\%$.

\subsection{Astrophysical factor of the DM annihilation}

The annihilation signal of DM particles relies on two factors: the
particle physical factor $W(E)$ (energy dependent) depending on the
particle property of DM, and the astrophysical factor $J(\psi)$
(spatial dependent) depending on the spatial distribution of DM. The
neutrino flux observed on the Earth (applicable also for $\gamma$)
can be written as
\begin{eqnarray}
\phi(E,\psi)&=&C\times W(E)\times J(\psi) \nonumber\\
            &=&\rho_{\odot}^2R_{\odot}\times\frac{1}{4\pi}\frac{\langle
\sigma v\rangle}{2m_{\chi}^2}\frac{{\rm d}N}{{\rm d}E}\times
\frac{1}{\rho_{\odot}^2R_{\odot}}\int_{LOS}\rho^2(l){\rm d}l,
\label{jpsi}
\end{eqnarray}
where $\rho_{\odot}=0.34$ GeV cm$^{-3}$ is the local DM density and
$R_{\odot}=8.5$ kpc is the distance of the Sun from the GC, $\psi$
is defined as the angle between the observational direction and the
GC direction relative to the observer, $\langle\sigma v\rangle$ is
the average value of annihilation cross section times relative
velocity, $m_{\chi}$ is the mass of DM particle, ${\rm d}N/{\rm d}E$
is the production spectrum of  $\nu$ per annihilation. The integral
path in Eq. (\ref{jpsi}) is along the line-of-sight (LOS). To
account for the contribution of substructures, we just need to
replace $\rho^2$ in Eq. (\ref{jpsi}) by
$\langle\rho^2\rangle=\rho_{smooth}^2+ \langle\rho_{sub}^2\rangle$
with \cite{yuan07}
\begin{equation}
\langle\rho_{sub}^2\rangle=\int {\rm d}M_{sub}\frac{{\rm d}N}
{{\rm d}M_{sub}\cdot4\pi r^2{\rm d}r}\left(\int_{V_{sub}}\rho_{sub}^2
{\rm d}V\right).
\label{averrho}
\end{equation}

\begin{figure}[!htb]
\begin{center}
\scalebox{0.7}{\includegraphics{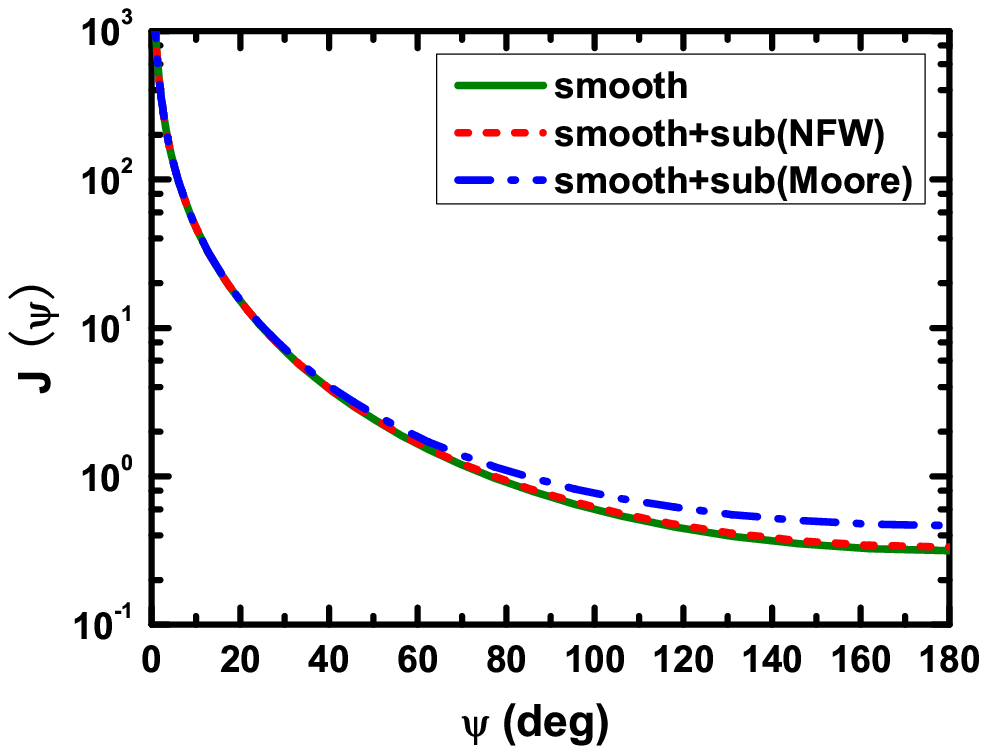}}
\scalebox{0.7}{\includegraphics{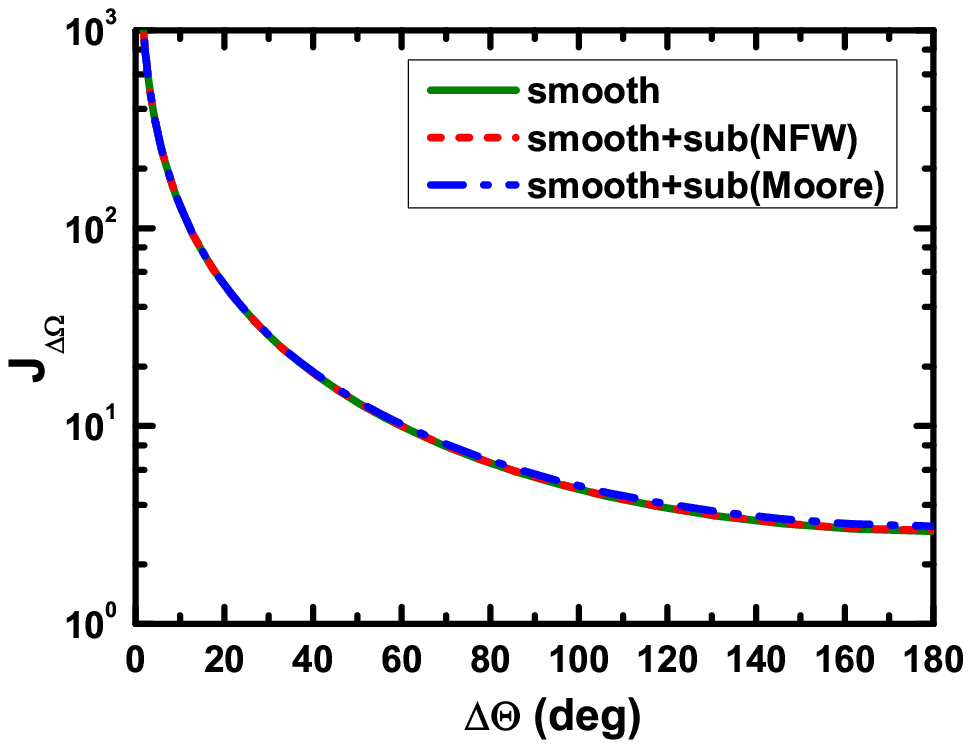}}
\scalebox{0.7}{\includegraphics{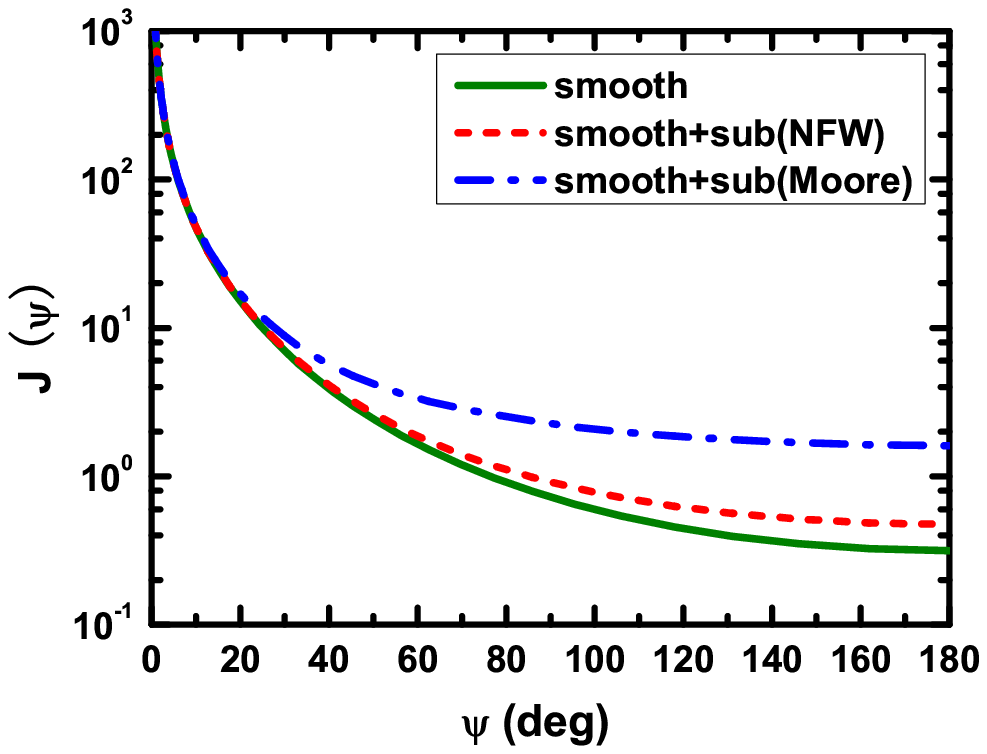}}
\scalebox{0.7}{\includegraphics{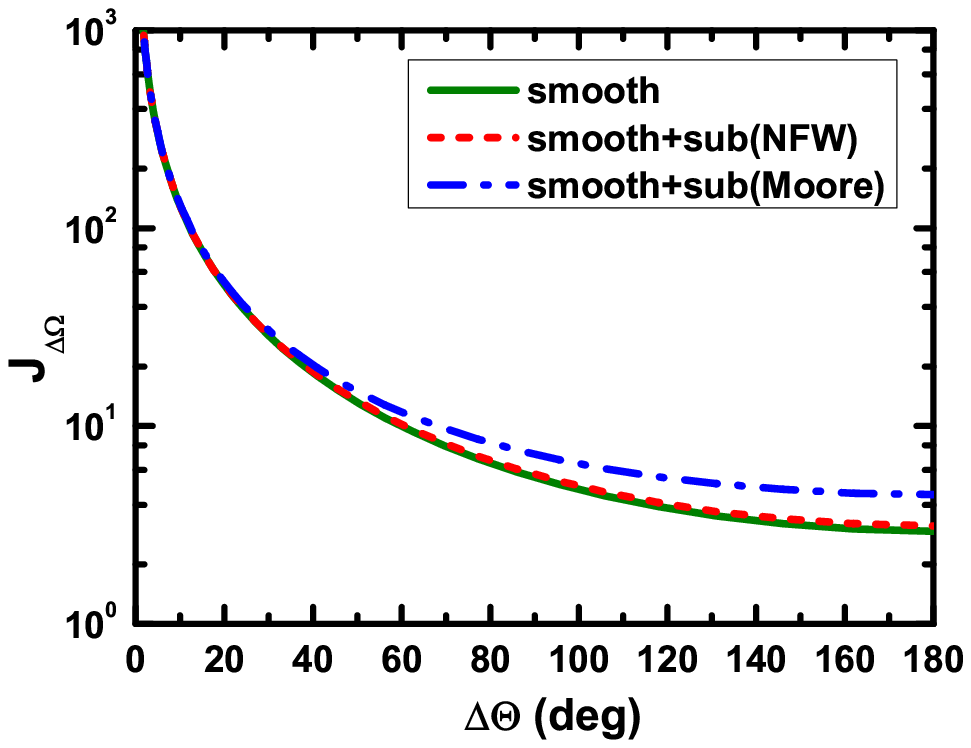}}
\scalebox{0.7}{\includegraphics{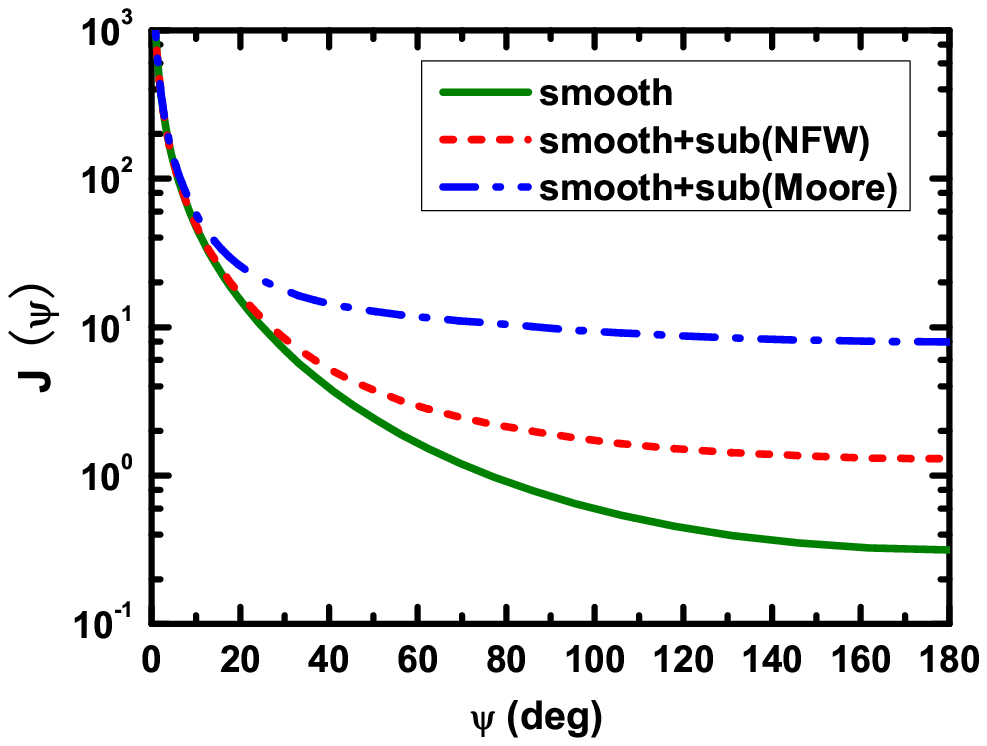}}
\scalebox{0.7}{\includegraphics{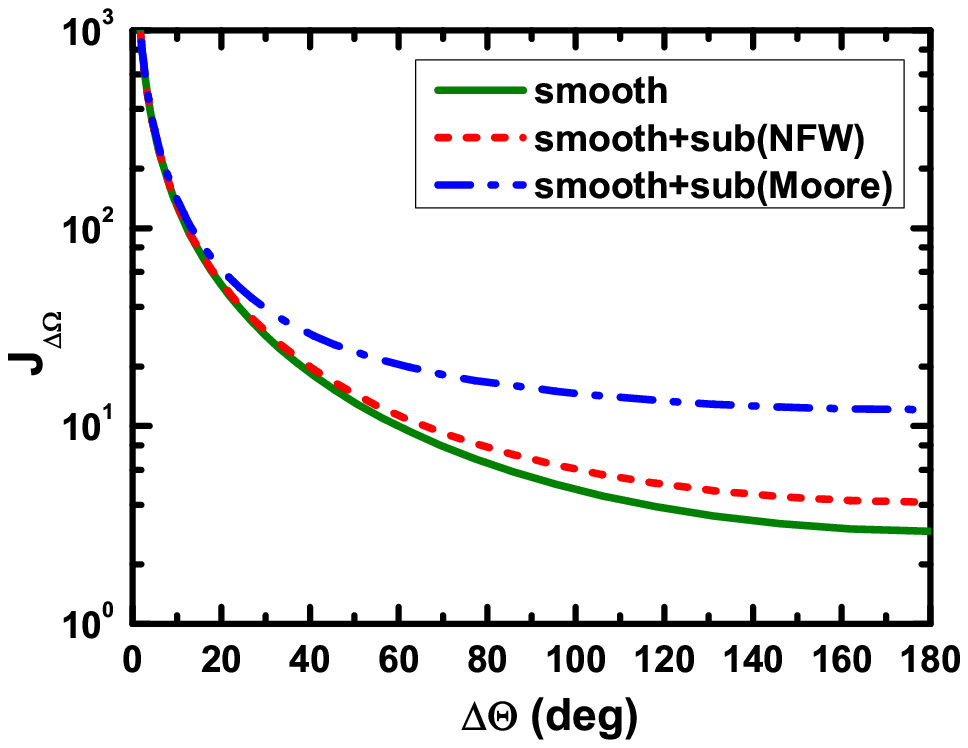}}
\caption{$J(\psi)$ as a function of $\psi$ (left column) and
$J_{\Delta\Omega}$ as a function of smooth angle
$\Delta\Theta$---half angle of the smooth cone centered at the
direction of GC (right column). In each figure the smooth (solid),
smooth+sub(NFW) (dashed) and smooth+sub(Moore) (dotted) are shown.
For each column, from top to bottom, the figures correspond to the
concentration models as ENS01, B01 and B01 multiplying by a factor
of 2, respectively.} \label{jpsieps}
\end{center}
\end{figure}

The average astrophysical factor within a solid angle $\Delta\Omega$
(e.g., the angular resolution of the detector) is defined as
\begin{equation}
J_{\Delta\Omega}=\frac{1}{\Delta\Omega}\int_{\Delta\Omega}J(\psi){\rm
d}\Omega .
\end{equation}
where $ J(\psi )$ is defined in Eq. (\ref{jpsi}). In Fig.
\ref{jpsieps} we show $J(\psi)$ as a function of $\psi$ and
$J_{\Delta\Omega}$ as a function of smooth angle $\Delta\Theta$,
which is the half angle of the cone centered at the direction of the
GC. Here $\Delta\Omega=2\pi(1-\cos\Delta\Theta)$. From the figures
we can see that the enhancement on $J(\psi)$ by substructures is
mainly at large angle, i.e., the direction deviation from the GC.
The contribution from sub-halos with Moore profile is about 8 times
larger than that of NFW profile (slightly different between various
concentration models). For ENS01 model the enhancement upon the
smooth component is very weak, while for the best combination
B01$\times 2$+Moore the boost factor can be as large as $\sim25$ at
the anti-GC direction. However, the enhancement of
$J_{\Delta\Omega}$ is not remarkable. For the halo-average
($\Delta\Theta=180^{\circ}$) the best case gives $\sim 4$ times
boost as shown in the right bottom figure in Fig. \ref{jpsieps}.

\subsection{DM substructure as point-like source}

In the previous section the contribution from substructures is
averaged over the whole MW (see Eq. (\ref{averrho})). It should be
noted that for the few massive sub-halos this average is
unreasonable. In this case, the massive sub-halo should be more
appropriately treated as point-like source and may be identified by
high angular resolution detector such as the IceCube. This feature
can also be utilized to suppress the background.

\begin{figure}[!htb]
\begin{center}
\includegraphics{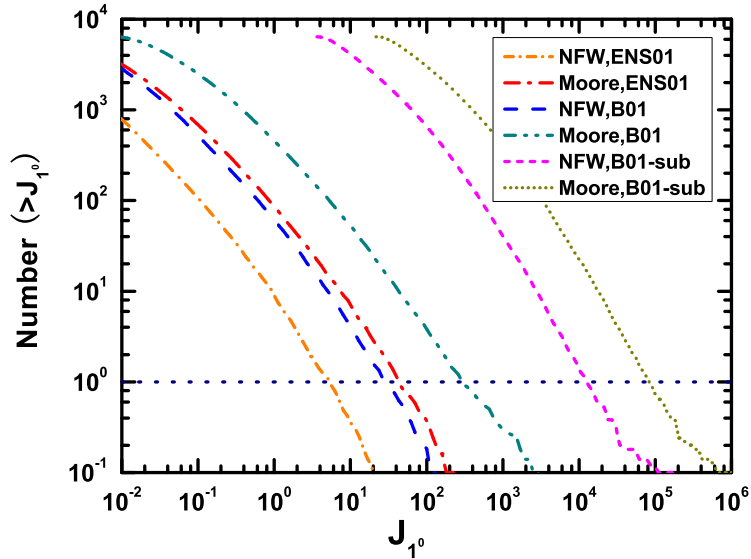}
\caption{Cumulative number of sub-halos whose astrophysical factor
$>J_{1^{\circ}}$ as a function of $J_{1^{\circ}}$. The dashed
horizon line corresponds to the case that only one subhalo will be
observed. }\label{jpsisubeps}
\end{center}
\end{figure}

In our work, a Monte-Carlo method is adopted to produce sub-halos
with mass larger than $10^6$ M$_{\odot}$ according to the
distribution function Eq. (\ref{number}). For this distribution
function, we find $N(>10^6$ M$_{\odot})\approx6400$. In our
numerical simulation, totally 50 MWs are generated, and the ``tidal
approximation'' is adopted. For each sub-halo, we calculate the
value of $J(\psi)$ and average within the cone with half angle equal
to $1^{\circ}$. Then we count the cumulative number of sub-halos
with astrophysical factor larger than a specified value, as shown in
Fig. \ref{jpsisubeps}. The concentration models for ENS01, B01 and
B01-subhalo (see Fig. \ref{cveps}) are adopted. We can see that the
ENS01 model gives the smallest astrophysical factor, while the
result from B01-subhalo model is larger by orders of magnitude than
other models. From the figure we can see that the uncertainties of
density distribution in sub-halos are very large. Given the particle
factor of DM annihilation, we can get the number of sub-halos with
flux higher than a specified value, such as the sensitivity of the
detector. On the contrary, if no source is seen, it puts a
constraint on the DM annihilation cross section \cite{bi07}.

\section{Constraint on dark matter
annihilation cross section from measurements of atmospheric neutrino
flux}

The atmospheric neutrino comes from hadronic interaction between
cosmic ray and atmosphere around the Earth. The observed neutrino
flux is consistent with theoretical prediction within the
uncertainties \cite{Ahrens:2002gq,Ashie:2005ik}. The neutrinos from
the DM annihilation should not be larger than the measured flux of
the neutrino. Thus this requirement gives bounds on the DM
annihilation cross section provided that the astrophysical factor,
as discussed in the above section, is known.  Actually in the
literature the constraints on the neutrino fluxes from cosmic
diffuse processes and the whole MW halo have been investigated in
Ref. \cite{Beacom:2006tt,Yuksel:2007ac}. In this work, influence on
neutrino flux by DM substructures will be scrutinized.

We assume that DM particles annihilate into pairs of neutrinos
following Ref. \cite{Beacom:2006tt}. The spectrum of the neutrinos
per flavor is a monochromatic line with $ dN_\nu  /dE_\nu   =
\frac{2}{3}\delta (E_\nu   - m_\chi  )$, where $ m_\chi$ is the mass
of the DM particle. All neutrino flavors are equally populated and
the neutrino and anti-neutrino are added together. We require that
the average neutrino flux from DM annihilation should not exceed the
average atmospheric neutrino flux, which is taken from
\cite{Honda:2006qj}.

We first consider the neutrino flux averaged over all directions
with smooth and sub-halo contributions. As shown by the left column
in Fig. \ref{jpsieps}, the sub-halo contribution to the neutrino
flux is insensitive to the direction \cite{bi06a}, while the smooth
contribution can vary by several orders of magnitude for GC and
anti-GC directions. The sub-halo contribution in anti-GC direction
is larger than that of the smooth one, but smaller in GC direction.
In other words, the neutrinos from anti-GC direction are mainly from
sub-halo DM annihilation.

Second, we investigate the effects on the neutrino flux from DM
sub-halo contribution. Massive sub-halo is more appropriately
treated as point-like source, as discussed above. We take smaller
angle cone to calculate the neutrino flux from such single sub-halo
and the atmospheric neutrino separately. The average neutrino flux
from sub-halo in this cone is required  not to exceed the
atmospheric neutrino flux. Combined with the input astrophysical
factor, we can set bounds on the DM annihilation cross section. This
approach is obviously based on the excellent angular resolution of
the neutrino telescope. The angular resolution of Super-Kamiokande
is $ \delta \theta (E) \simeq 30^ \circ   \times \sqrt {GeV/E}$
\cite{Ashie:2005ik}. For  $ E
> 100GeV$ neutrino, the angular resolution can reach about $ 3^ \circ$ for
Super-Kamiokande, while $1^ \circ$ for IceCube \cite{Ahrens:2003ix}.
In our evaluation we adopt a $ 1^ \circ$ half-angle cone for
IceCube. Note that for IceCube, the candidate source should be in
the northern sky, and its threshold energy is $ 50GeV$
\cite{Ahrens:2003ix}.

In order to obtain the constraint on the DM annihilation cross
section, we must input the astrophysical factor. We use average
neutrino flux in a cone $ \phi _{\Delta \Omega }$ by simply
replacing $ J(\psi )$ with $ J_{\Delta \Omega }$ in Eq.
(\ref{jpsi}). Since the DM induced neutrino is sharply peaked, we
average neutrino flux within the energy bin around $ E_\nu = m_\chi$
to compare with the atmospheric neutrinos. And the energy bin width
is taken as $ \Delta \log _{10} E = 0.3$ within the energy
resolution limits of the neutrino detectors Super-Kamiokande
\cite{Desai:2004pq,Ashie:2005ik}and IceCube \cite{Ahrens:2003ix}.
The $ J_{\Delta \Omega }$ used in sub-halo point source is taken
from Fig. \ref{jpsisubeps} when cumulative number equals to one.
Based on the astrophysical factor input this way, the derived
annihilation cross section gives the sensitivity that IceCube may
find at least one sub-halo. $ J_{\Delta \Omega }$ for different
sub-halo profiles and concentration models are given in Table.
\ref{table1}.

\begin{table}[htb]
\begin{tabular}{c|cc|ccc}
\hline \hline
  & \multicolumn{2}{c|}{Halo average} & \multicolumn{3}{c}{Point-like} \\
\hline
  & smooth & smooth + subhalo & ENS01 & B01 & B01-subhalo \\
 \hline
  NFW & 3.0 & 4.1 & 5.3 & 28.9  & 12763.7\\
 \hline
  Moore & --- & 12.1 & 43.3 & 276.3 & 82110.2\\
  \hline
  \hline
\end{tabular}
\caption{$ J_{\Delta \Omega }$ for different sub-halo profiles and
concentration models. The values listed here correspond to the
cumulative number equal to $1$ as shown in Fig. \ref{jpsisubeps}.
The smooth contribution to $ J_{\Delta \Omega }$ is fixed to be NFW
profile. 'Halo average' means averaging contributions over the whole
MW halo. For the sub-halo contributions  we choose the concentration
model of B01$\times$2, as shown in the bottom-right figure in Fig.
\ref{jpsieps}. ENS01, B01 and B01subhalo represent different
concentration models when considering DM substructure as point-like
sources. For point-like sub-halos, $ J_{\Delta \Omega }$ is averaged
in a $ 1^ \circ$ half-angle cone around the center of the sub-halo.
}\label{table1}
\end{table}

From Table \ref{table1} we can see that $ J_{\Delta \Omega }$ is 3.0
for the smooth case. The sub-halo contribution with Moore profile is
9.1, while only 1.1 for NFW profile. If one averages the
contributions over the whole sky, the sub-halo contribution can
enhance the neutrino flux slightly. As shown in Fig. \ref{jpsieps},
the enhancement of sub-halo is large at the anti-GC direction which
can reach about $ 100 \sim 1000$, but the enhancement at the GC
direction is small. Note that the flux from the GC is much larger
than that from the anti-GC, it is natural to expect the {\em
averaging} sub-halo contribution should not be significant.
Therefore the point-like sub-halo should be naturally more
important, as shown in Table \ref{table1}. For the different
concentration model ENS01, B01 and B01subhalo, $ J_{\Delta \Omega }$
is much larger than the average contribution, though the uncertainty
is very large. The numbers in different concentration models can be
understood from Fig. \ref{cveps}, which indicates that the large $
c_v$ corresponds to the large $ J_{\Delta \Omega }$. Virial radius $
r_v$ defined in Eq. \ref{cv} is the boundary that gravitational
force can sustain itself, and it is only related to virial mass $
M_v$ but not the mass distribution. The density profiles with NFW or
Moore distribution behave both like $ r^{ - 3}$ at large radii and $
r^{ - 1}$ or $ r^{ - 1.5}$ at small radii separately. The radius $
r_{ - 2}$ is a measure of the density profile $ \rho (r)$, namely
the dark matter  distributes mainly within the region $ r < r_{ -
2}$. Thus larger $ c_v$ means more dark matter concentrating in the
sub-halo center, which results in a larger $ J_{\Delta \Omega }$.

\begin{figure}[!htb]
\begin{center}
\includegraphics{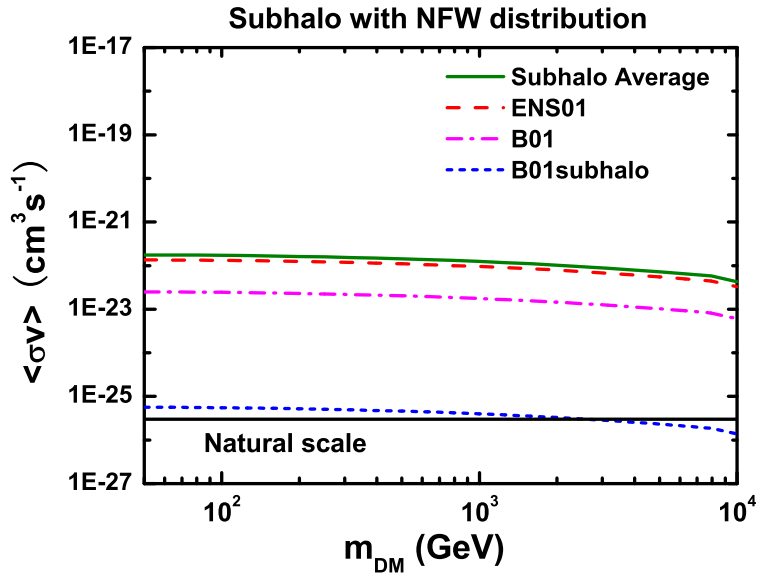}
 \caption{The upper
bounds on the DM annihilation cross section $\langle\sigma v\rangle$
as a function of dark matter mass with NFW subhalo profile. The
upper solid lines represent the constraints from $ 180^ \circ$
half-angle cone average. The other lines represent constraints from
sub-halo point source with $ 1^ \circ$ resolution from the IceCube.
The dashed, dash-dotted and short-dashed lines are plotted for
different concentration models. The lower solid line corresponds to
$\langle\sigma v\rangle\sim3\times10^{-26}cm^3s^{-1}$ for the
natural scale to produce the correct thermal relic
density.}\label{sigmavn}
\end{center}
\end{figure}

\begin{figure}[!htb]
\begin{center}
\includegraphics{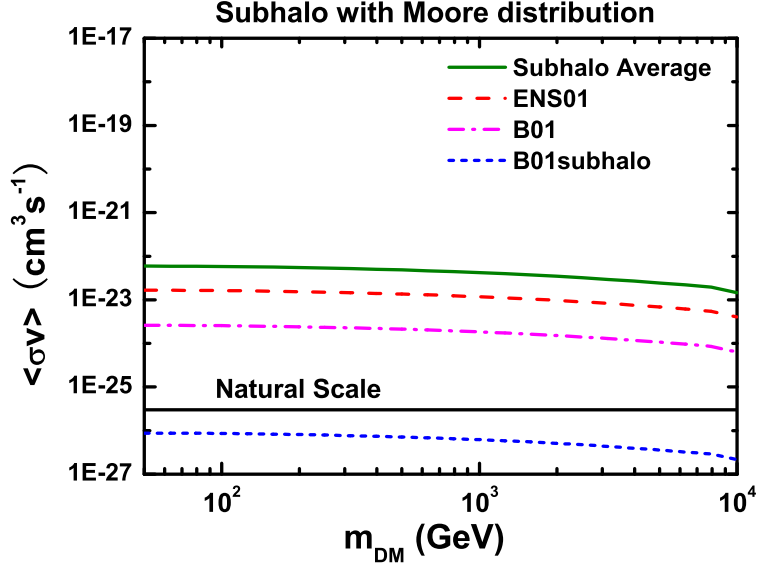}
 \caption{Same with Fig. \ref{sigmavn} but with
Moore subhalo profile. }\label{sigmavm}
\end{center}
\end{figure}

Once the astrophysical factor is known, bounds on the dark matter
annihilation cross section $\langle\sigma v\rangle$ can be derived
by requiring the DM induced neutrino flux less than the measured
ones. In Fig. \ref{sigmavn} and Fig. \ref{sigmavm}, the upper bounds
on $\langle\sigma v\rangle$  are plotted for two DM profiles. Also
shown in the figures is $\langle\sigma v\rangle$ for the natural
scale of dark matter. The upper solid line correspond to the
constraints from $ 180^ \circ$ half-angle cone averaging
contributions. The other lines represent the sub-halo point source
with $ 1^ \circ$ resolution from the IceCube. The dashed,
dash-dotted and short-dashed lines are plotted with different
concentration models. The constraints from the whole halo average
are slightly improved compared with the smooth case in Refs.
\cite{Beacom:2006tt,Yuksel:2007ac}. For the sub-halo as the point
source, the constraints are significantly improved. Moreover the
upper bounds of $\langle\sigma v\rangle$ are about $10^{-23}\sim
10^{-24}$ cm$^3$ s$^{-1}$ for B01 model, and can even reach
$\sim10^{-26}$ cm$^3$ s$^{-1}$ for B01-subhalo model. Such bound is
even lower than $\langle\sigma v\rangle$ for the natural scale which
can induce the correct relic density of DM.

\section{Conclusions and discussions}

In this paper, by requiring the DM induced neutrino flux less than
the measured ones, we give the improved upper bounds on the DM
annihilation cross section $\langle\sigma v\rangle$ with the DM
substructure effects included. Here we assume the DM particles
annihilate into neutrinos solely following the previous works. The
observed neutrino flux depends on the particle physical and
astrophysical factors. Thus we first investigate the astrophysical
factor. Several different DM profiles and sub-halo concentration
models are adopted based on the numerical simulations. Our studies
show that at the anti-GC direction, the enhancement factor of
sub-halos for B01$\times2$ model is about 3 and  25 for NFW and
Moore profiles respectively. While the whole-sky average ($halo\
average$, with cone half-angle $180^{\circ}$) does not have
prominent enhancement. The best case of our adopted models is only
$\sim4$ times larger than the smooth ones (see bottom-right figure
in Fig. \ref{jpsieps}). If we take the $30^{\circ}$ angular average
($halo\ angular$ in Ref. \cite{Yuksel:2007ac}), there is almost no
enhancement, as can be seen in Fig. \ref{jpsieps}. This is because
the enhancement from sub-halos is spatially dependent on the MW
halo, instead of a universal one \cite{bi06b}. On one hand, the
smooth component increases more rapidly and dominates the
annihilation flux near the GC; on the other hand, the tidal
disruption on sub-halos is most effective close to the GC. Thus the
effect of substructures is not significant near the GC. In the cases
of $halo\ average$ and $halo\ angular$, the GC contributions are
included and play a dominant role in the total flux, therefore no
remarkable enhancement from sub-halos is found.

In this paper we emphasize the important role of the massive
sub-halos (e.g., $M_{sub}>10^6$M$_{\odot}$). Since the number of
massive sub-halo is small, it should be regarded as the point-like
source. For the massive sub-halos, the high angular resolution of
neutrino detector can be utilized to suppress the atmospheric
neutrino background, thus the constraints on $\langle\sigma
v\rangle$ are expected to be improved. The angular resolution $\sim
1^{\circ}$ for energy greater than $50$ GeV of the forthcoming
experiment IceCube is employed \footnote{ For Super-Kamiokande, the
resolution angle is $\sim 3^{\circ}$ for $E>100$ GeV and the
background is $9$ times greater.}. The neutrino signal flux in a
cone with half-angle $1^{\circ}$ is calculated and compared with the
atmospheric background in the same cone. We found that the
constraints on $\langle\sigma v\rangle$ are indeed improved
significantly. Note that the constraints are model-dependent. For
the moderate case B01+NFW, we find the upper bound of $\langle\sigma
v\rangle$ is about $10^{-23}$ cm$^3$ s$^{-1}$. While for the
concentration model B01-subhalo, the bound can reach $10^{-26}$
cm$^3$ s$^{-1}$, which is even lower than $\langle\sigma v\rangle$
for the natural scale which can induce the correct relic density of
DM.

It should be noted that DM can annihilate into final states other
than neutrinos. Thus the assumption that DM annihilates into only
neutrinos gives the most conservative bound on the DM annihilation
cross section and is independent of the particle properties of DM
particle.

Neutrinos are  thought to be an important complementary particles
for DM indirect searches besides  $\gamma$-rays and charged
anti-particles. It is shown that the detectability of $\gamma$-rays
from DM sub-halos on GLAST is optimistic \cite{kuhlen08}. The
effects of subhalos on positrons \cite{cumber07, lavalle07} and
antiprotons \cite{lavalle08} are also investigated. The combination
and cross check of different kinds of signals will be very crucial
to identify the DM sources and investigate the properties of DM
particles.

\section{ Acknowledgements}

This work was supported in part by the Natural Sciences Foundation
of China (Nos. 10775001, 10635030, 10575111, 10773011),
by the trans-century fund of Chinese Ministry of Education,
and by the Chinese Academy of Sciences under the grant No. KJCX3-SYW-N2.


\begin{thebibliography}{99}

\bibitem{Jungman:1995df}
  G.~Jungman, M.~Kamionkowski and K.~Griest,
  Phys.\ Rept.\  {\bf 267}, 195 (1996).

\bibitem{Begeman:1991iy}
  K.~G.~Begeman, A.~H.~Broeils and R.~H.~Sanders,
  Mon.\ Not.\ Roy.\ Astron.\ Soc.\  {\bf 249}, 523 (1991).

\bibitem{Persic:1995ru}
  M.~Persic, P.~Salucci and F.~Stel,
  Mon.\ Not.\ Roy.\ Astron.\ Soc.\  {\bf 281}, 27 (1996).

\bibitem{Tyson95} J. A. Tyson and P. Fischer, Astrophys. J. {\bf 446} 55
  (1995).

\bibitem{White93}
  S.~D.~M.~White, J.~F.~Navarro, A.~E.~Evrard and C.~S.~Frenk,
  Nature {\bf 366}, 429 (1993).

\bibitem{Peebles71}
  P.~J.~E.~Peebles, {\it Physical Cosmology} (Princeton: Princeton
  University Press) (1971).


\bibitem{Spergel03}
  D.~N.~Spergel {\it et al.}  [WMAP Collaboration],
  Astrophys.\ J.\ Suppl.\  {\bf 148}, 175 (2003).

\bibitem{Hinshaw:2008kr}
  G.~Hinshaw {\it et al.}  [WMAP Collaboration],
  arXiv:0803.0732 [astro-ph].

\bibitem{Bertone04}
  G.~Bertone, D.~Hooper and J.~Silk,
  Phys.\ Rept.\  {\bf 405}, 279 (2005).

\bibitem{Liu:2008kz}
  J.~Liu, P.~f.~Yin and S.~h.~Zhu,
  Phys.\ Rev.\  D {\bf 77}, 115014 (2008) and references therein.

\bibitem{Gondolo:1999ef}
  P.~Gondolo and J.~Silk,
  Phys.\ Rev.\ Lett.\  {\bf 83}, 1719 (1999).

\bibitem{bi06a}
  X.~J.~Bi,
  Nucl.\ Phys.\  B {\bf 741}, 83 (2006).

\bibitem{bi06b}
  X.~J.~Bi, J.~Zhang, Q.~Yuan, J.~L.~Zhang and H.~Zhao,
  arXiv:astro-ph/0611783.

\bibitem{yuan07}
  Q.~Yuan and X.~J.~Bi,
  JCAP {\bf 0705}, 001 (2007).

\bibitem{Desai:2004pq}
  S.~Desai {\it et al.}  [Super-Kamiokande Collaboration],
  Phys.\ Rev.\  D {\bf 70}, 083523 (2004)
  [Erratum-ibid.\  D {\bf 70}, 109901 (2004)].

\bibitem{Ahrens:2003fg}
  J.~Ahrens {\it et al.}  [AMANDA Collaboration],
  Nucl.\ Instrum.\ Meth.\  A {\bf 524} (2004) 169;
  M.~Ackermann {\it et al.}  [The AMANDA Collaboration],
  Phys.\ Rev.\  D {\bf 71}, 077102 (2005).

\bibitem{Beacom:2006tt}
  J.~F.~Beacom, N.~F.~Bell and G.~D.~Mack,
  Phys.\ Rev.\ Lett.\  {\bf 99}, 231301 (2007).

\bibitem{Yuksel:2007ac}
  H.~Yuksel, S.~Horiuchi, J.~F.~Beacom and S.~Ando,
  Phys.\ Rev.\  D {\bf 76}, 123506 (2007).

\bibitem{Kachelriess:2007aj}
  M.~Kachelriess and P.~D.~Serpico,
  Phys.\ Rev.\  D {\bf 76}, 063516 (2007).

\bibitem{Bell:2008ey}
  N.~F.~Bell, J.~B.~Dent, T.~D.~Jacques and T.~J.~Weiler,
  arXiv:0805.3423 [hep-ph].

\bibitem{Dent:2008qy}
  J.~B.~Dent, R.~J.~Scherrer and T.~J.~Weiler,
  arXiv:0806.0370 [astro-ph].

\bibitem{Pradier:2008iv}
  T.~Pradier and f.~t.~A.~Collaboration,
  arXiv:0805.2545 [astro-ph].

\bibitem{Ahrens:2003ix}
  J.~Ahrens {\it et al.}  [IceCube Collaboration],
  Astropart.\ Phys.\  {\bf 20}, 507 (2004).

\bibitem{nfw97}
  J.~F.~Navarro, C.~S.~Frenk and S.~D.~M.~White,
  Astrophys.\ J.\  {\bf 490}, 493 (1997).

\bibitem{moore98}
  B.~Moore, F.~Governato, T.~Quinn, J.~Stadel and G.~Lake,
  Astrophys.\ J.\  {\bf 499}, L5 (1998).

\bibitem{burkert95}
  A.~Burkert,
  IAU Symp.\  {\bf 171}, 175 (1996)
  [Astrophys.\ J.\  {\bf 447}, L25 (1995)].

\bibitem{Salucci:2000ps}
  P.~Salucci and A.~Burkert,
  Astrophys.\ J.\  {\bf 537}, L9 (2000).

\bibitem{diemand05}
  J.~Diemand, B.~Moore and J.~Stadel,
  Nature {\bf 433}, 389 (2005).

\bibitem{reed05}
  D.~Reed {\it et al.},
  Mon.\ Not.\ Roy.\ Astron.\ Soc.\  {\bf 357}, 82 (2005).

\bibitem{berezinsky92}
  V.~S.~Berezinsky, A.~V.~Gurevich and K.~P.~Zybin,
  Phys.\ Lett.\  B {\bf 294}, 221 (1992).

\bibitem{lavalle08}
  J.~Lavalle, Q.~Yuan, D.~Maurin and X.~J.~Bi,
  Astron.\ Astrophys.\  {\bf 479}, 427 (2008).

\bibitem{bullock01}
  J.~S.~Bullock {\it et al.},
  Mon.\ Not.\ Roy.\ Astron.\ Soc.\  {\bf 321}, 559 (2001).

\bibitem{eke01}
  V.~R.~Eke, J.~F.~Navarro and M.~Steinmetz,
  Astrophys.\ J.\  {\bf 554}, 114 (2001).

\bibitem{buote07}
  D.~A.~Buote, F.~Gastaldello, P.~J.~Humphrey, L.~Zappacosta, J.~S.~Bullock, F.~Brighenti and W.~G.~Mathews,
  Astrophys.\ J.\  {\bf 664}, 123 (2007).

\bibitem{comerford07}
  J.~M.~Comerford and P.~Natarajan,
  Mon.\ Not.\ Roy.\ Astron.\ Soc.\  {\bf 379}, 190 (2007).

\bibitem{cola06}
  S.~Colafrancesco, S.~Profumo and P.~Ullio,
  Astron.\ Astrophys.\  {\bf 455}, 21 (2006).

\bibitem{wilkinson99}
  M.~I.~Wilkinson and N.~W.~Evans,
  Mon.\ Not.\ Roy.\ Astron.\ Soc.\  {\bf 310}, 645 (1999).

\bibitem{sakamoto03}
  T.~Sakamoto, M.~Chiba and T.~C.~Beers,
  Astron.\ Astrophys.\  {\bf 397}, 899 (2003).

\bibitem{smith07}
  M.~C.~Smith {\it et al.},
  Mon.\ Not.\ Roy.\ Astron.\ Soc.\  {\bf 379}, 755 (2007).

\bibitem{xue08}
  X.~X.~Xue {\it et al.},
  arXiv:0801.1232 [astro-ph].

\bibitem{tormen98}
  G.~Tormen, A.~Diaferio and D.~Syer, Mon.\ Not.\ Roy.\ Astron.\ Soc.\ {\bf
  299}, 728 (1998).

\bibitem{klypin99}
  A.~A.~Klypin, S.~Gottlober and A.~V.~Kravtsov,
  Astrophys.\ J.\  {\bf 516}, 530 (1999).

\bibitem{moore99}
  B.~Moore, S.~Ghigna, F.~Governato, G.~Lake, T.~Quinn, J.~Stadel and P.~Tozzi,
  Astrophys.\ J.\  {\bf 524} (1999) L19.

\bibitem{ghigna00}
  S.~Ghigna, B.~Moore, F.~Governato, G.~Lake, T.~Quinn and J.~Stadel,
  Astrophys.\ J.\  {\bf 544}, 616 (2000).

\bibitem{springel01}
  V.~Springel, S.~D.~M.~White, G.~Tormen and G.~Kauffmann,
  Mon.\ Not.\ Roy.\ Astron.\ Soc.\  {\bf 328}, 726 (2001).

\bibitem{zentner03}
  A.~R.~Zentner and J.~S.~Bullock,
  Astrophys.\ J.\  {\bf 598}, 49 (2003).

\bibitem{lucia04}
  G.~De Lucia {\it et al.},
  Mon.\ Not.\ Roy.\ Astron.\ Soc.\  {\bf 348}, 333 (2004).

\bibitem{diemand04}
  J.~Diemand, B.~Moore and J.~Stadel,
  Mon.\ Not.\ Roy.\ Astron.\ Soc.\  {\bf 352}, 535 (2004).

\bibitem{helmi02}
  A.~Helmi, S.~D.~M.~White and V.~Springel,
  Phys.\ Rev.\  D {\bf 66}, 063502 (2002).

\bibitem{gao04}
  L.~Gao, S.~D.~M.~White, A.~Jenkins, F.~Stoehr and V.~Springel,
  Mon.\ Not.\ Roy.\ Astron.\ Soc.\  {\bf 355} (2004) 819.

\bibitem{shaw06}
  L.~Shaw, J.~Weller, J.~P.~Ostriker and P.~Bode,
  Astrophys.\ J.\  {\bf 646}, 815 (2006).

\bibitem{hofmann01}
  S.~Hofmann, D.~J.~Schwarz and H.~Stoecker,
  Phys.\ Rev.\  D {\bf 64}, 083507 (2001).

\bibitem{chen01}
  X.~l.~Chen, M.~Kamionkowski and X.~m.~Zhang,
  Phys.\ Rev.\  D {\bf 64}, 021302 (2001).

\bibitem{green05}
  A.~M.~Green, S.~Hofmann and D.~J.~Schwarz,
  JCAP {\bf 0508}, 003 (2005).

\bibitem{bi07}
  X.~J.~Bi,
  Phys.\ Rev.\  D {\bf 76}, 123511 (2007).

\bibitem{Ahrens:2002gq}
  J.~Ahrens {\it et al.}  [AMANDA Collaboration],
  Phys.\ Rev.\  D {\bf 66}, 012005 (2002).

\bibitem{Ashie:2005ik}
  Y.~Ashie {\it et al.}  [Super-Kamiokande Collaboration],
  Phys.\ Rev.\  D {\bf 71}, 112005 (2005).

\bibitem{Honda:2006qj}
  M.~Honda, T.~Kajita, K.~Kasahara, S.~Midorikawa and T.~Sanuki,
  Phys.\ Rev.\  D {\bf 75}, 043006 (2007).

\bibitem{kuhlen08}
  M.~Kuhlen, J.~Diemand and P.~Madau,
  arXiv:0805.4416 [astro-ph].

\bibitem{cumber07}
  D.~T.~Cumberbatch and J.~Silk,
  Mon.\ Not.\ Roy.\ Astron.\ Soc.\  {\bf 374}, 455 (2007).

\bibitem{lavalle07}
  J.~Lavalle, J.~Pochon, P.~Salati and R.~Taillet,
  Astron.\ Astrophys.\  {\bf 462}, 827 (2007).


\end{thebibliography}
\end{document}